\def\hcorrection#1{\advance\hoffset by #1 }
\def\vcorrection#1{\advance\voffset by #1 }

\documentstyle[11pt]{article}

\vcorrection{-0.70in}
\hcorrection{-0.40in}
\begin{document}

\vspace{0.3in}

\begin{center}
{\Large
\bf Analytical solutions of the lattice Boltzmann BGK model}
\end{center}

\vspace{0.3in}

\begin{center}
Qisu Zou$^{1,2}$,
Shuling Hou$^{1,3}$,
Gary D. Doolen$^{1}$,
\end{center}

\vspace{0.3in}

${}^{1}${\footnotesize  Center for Nonlinear Studies and Theoretical
Division, Los Alamos National Laboratory,} \\
\hspace*{0.3in}{\footnotesize Los Alamos, NM 87545}

${}^{2}${\footnotesize  Department of
Mathematics, Kansas State University, Manhattan, KS 66506.}

${}^{3}${\footnotesize  Department of Mechanical Engineering,
Kansas State University, Manhattan, KS 66506}

\begin{center}
{\bf Abstract}
\end{center}

Analytical solutions of the two dimensional triangular and square
lattice Boltzmann
BGK models
have been obtained for the plain Poiseuille flow and
the plain Couette  flow.  The analytical solutions are written in terms
of the characteristic velocity of the flow, the single relaxation time
 $\tau $
and the lattice spacing. The analytic solutions  are the exact
representation of these two flows without any approximation.

\section{Introduction}

 Since the appearance of  lattice gas automata (LGA) and its later
derivative, the lattice Boltzmann
equation method (LBE) as alternative computational methods to study transport
phenomena, some analytical solutions have been obtained for
these methods for nonuniform flows in 2D or 3D models
\cite{hen,corn,luo,ginz}.
They are based on linearized Boltzmann models,
and they have employed approximations.  In \cite{hen,corn,luo},
the solution around a global equilibrium with constant density and
isotropic velocity (zero velocity) was considered. In \cite{ginz}, the
first order and second order of deviation of distribution function  from
equilibrium were assumed to take  a certain form in terms of flow quantities,
and the coefficients in this form were obtained using a
Chapman-Enskog procedure.
These analytic results provided insight for applications of the methods.
For example, the analytical results allow one
to calculate viscosity from given
collision rules and to estimate and to improve bounce-back boundary conditions.
They are valuable in enhancing our understanding of the method.
Nevertheless, analytical solutions for real flows with
boundaries like the Poiseuille flow, which is represented exactly
by a second-order finite-difference scheme on a uniform mesh,
have not been obtained previously for LGA or LBE.
One reason may be that the boundary conditions used in LGA and LBE
are not exact.
For example, bounce-back or a combination of bounce-back and specular
reflection  \cite{corn} for modeling the non-slip   boundary condition are
only approximate. The effective non-slip boundary is inside
the bounce-back row \cite{corn,ginz}.
Recently, Noble {\em et al.} \cite{nob} have proposed a boundary condition
for the lattice Boltzmann BGK model (LBGK) on a triangular lattice.
When this boundary condition was applied to plain Poiseuille flow, the
steady state solution of the distribution function gave a parabolic velocity
profile up to machine accuracy.  The result suggests the existence of
an analytical solution to LBGK, which is an exact representation of
the Poiseuille flow.
This analytical solution is derived in this paper together with
an analytical solution for the plain Couette flow.

\section{Analytical Solutions of the triangular lattice LBGK model}

First let us consider the lattice Boltzmann model on a triangular lattice.
For a channel
flow, a triangular lattice is constructed as shown in Fig.~1.
There are two types of particles on each node of an FHP model:
rest particles (type 0) with ${\bf e}_0 = 0$  and
moving particles (type 1 to 6) with unit velocity
${\bf e}_i = (\cos ((i-1) \pi/6), \sin ((i-1)\pi/6), \ i = 1, \cdots, 6$
along 6 directions.
Consider the particle distribution functions $f_i({\bf x}, t)$, which represent
the probability of finding a particle at node ${\bf x} = (x,y)$ and time
$t$ with velocity ${\bf e}_i$.
The lattice Boltzmann BGK model proposed in \cite{ccmm,qian2}
is the equation for the evolution of $f_i$:
\begin{equation}
 f_i({\bf x}+\delta {\bf e}_i,t+\delta  )
-f_i({\bf x},t)
=-\frac{1}{\tau} [f_i({\bf x},t)-f_i^{(0)}({\bf x},t)], \;\; i = 0, \cdots, 6
\label{eq:4}
\end{equation}
where $f_i^{(0)}({\bf x},t) $ is the equilibrium distribution of the
particle of type $i$ at
${\bf x}, t$, the  right hand side represents the collision term
 and $\tau $ is the single relaxation time which controls the rate
of approach to equilibrium.
The density per node, $ \rho $, and the macroscopic flow velocity, ${\bf u}$,
are defined in terms of the particle distribution function by
\begin{equation}
 \sum_{i=0}^6 f_i=\rho,  \ \ \ \
 \sum_{i=1}^6 f_i {\bf e}_i =\rho {\bf u}.
\label{eq:2}
\end{equation}
The equilibrium distribution functions depend only on local density and
velocity.
A suitable equilibrium distribution for the FHP model
 can be chosen in the following form  \cite{qian2}:
\begin{eqnarray}
 f_{0}^{(0)}=d_0-\rho u^2 = \alpha \rho -\rho u^2 ,  \mbox{\hspace{1.5in}}
\nonumber\\
 f_{i}^{(0)}=d+\frac{1}{3}\rho[({\bf e}_{i}\cdot {\bf u})+2({\bf e}_i
\cdot {\bf u})^2-\frac{1}{2}{\bf u} \cdot {\bf u}], \;\; i=1, \cdots, 6 ,
\label{eq:3}
\end{eqnarray}
where $\alpha$ is an adjustable parameter, $d=\frac{\rho-d_0}{6}$, and
$ \sum_{i} f_i^{(0)}=\rho, \; \sum_{i} f_i^{(0)} {\bf e}_i =\rho {\bf u}$.
Note that Eq.~(\ref{eq:4}) is written in physical units with
the value of the lattice link being $\delta$,
Using unit speed for particles (with some
physical time unit), a time step has the value of $\delta$ as well.
A Chapman-Enskog procedure can be applied to Eq.~(\ref{eq:4}) to derive
the macroscopic equations  of the model.  They are given by:
the continuity equation (with an error term $O(\delta^2)$ being omitted):
\begin{equation}
\frac{\partial \rho}{\partial t} + \nabla \cdot (\rho {\bf u})=0,
\label{eq:5}
\end{equation}
and the momentum equation (with terms of  $O(\delta^2)$ and $O(\delta u^3)$
 being omitted):
\begin{equation}
\partial_{t}(\rho u_{\alpha})+\partial_{\beta}(\rho u_{\alpha} u_{\beta})
=
-\partial_{\alpha}(c_s^2 \rho)+\partial_{\beta}(2 \nu \rho S_{\alpha \beta}),
\label{eq:6}
\end{equation}
where the Einstein  summation convention is used,
$S_{\alpha \beta}=\frac{1}{2}(\partial_{\alpha}u_{\beta}+\partial_{\beta}
u_{\alpha})$ is the strain-rate tensor, the pressure is given by
$p = c_s^2 \rho$, where $c_s$ is the speed of sound with
${\displaystyle c_s^2= \frac{1-\alpha}{2}, }$
and
${\displaystyle \nu=\frac{2 \tau -1}{8} \delta}$,
with $\nu $ being the  the kinematic viscosity.
The form of the error terms and the derivation of these equations can be found
in \cite{qian3,hou}.
The macroscopic equations of LBGK represent the incompressible Navier-Stokes
equations in the limit as $\delta \rightarrow 0$, $\rho \rightarrow \rho_0 $
(a constant )
and the Mach number approaches zero,

The plain  Poiseuille flow in a channel with width $2 L $ and
velocity ${\bf u} = (u,v) $  is given by:
\begin{equation}
 u = u_0 (1 - \frac{y^2}{L^2}) , \;\;\;\; v = 0, \;\;\;\;
 \frac{\partial p}{\partial x} = - G, \;\;\;\;
 \frac{\partial p}{\partial y} = 0, \;\;\;\;
\label{eq:8}
\end{equation}
where $G$ is a constant related to the characteristic velocity $u_0$ by
\begin{equation}
 G = 2 \rho \nu u_0/L^2 ,
\label{eq:9}
\end{equation}
  and the flow density $\rho$ is a constant.
This is an exact solution of the incompressible Navier-Stokes equations:
\begin{eqnarray}
 \nabla \cdot {\bf u} = 0, \mbox{\hspace{2in}} \nonumber \\
\partial_{t}(\rho u_{\alpha})+\partial_{\beta}(\rho u_{\alpha} u_{\beta})
=- \partial_{\alpha}p+ \mu \partial_{\beta \beta} u_{\alpha} ,
\label{eq:10}
\end{eqnarray}
where $\mu = \rho \nu$.
Without loss of generality, we assume that $L = 1$ (a simple scaling
of $y' = y/L$ makes $y' \in [-1, 1]$).
To approximate  Poiseuille flow using the lattice Boltzmann model,
it is convenient
to replace the constant gradient by a body force ${\bf g}$
so that $\rho {\bf g} = -\nabla p$.
The momentum equation of N-S equations with a body force is:
\begin{equation}
\partial_{t}(\rho u_{\alpha})+\partial_{\beta}(\rho u_{\alpha} u_{\beta})
=- \partial_{\alpha}p+ \rho g_{\alpha} +
 \mu \partial_{\beta \beta} u_{\alpha} .
\label{eq:11}
\end{equation}
The Poiseuille flow can be generated with a body force ${\bf g}$
with $ \rho g_x = G, g_y = 0$,
where the pressure is held constant.
An LBGK which incorporates the body force is a modification of
Eq.~(\ref{eq:4}) given by:
\begin{equation}
 f_i({\bf x}+\delta {\bf e}_i,t+\delta  )
-f_i({\bf x},t)
=-\frac{1}{\tau} [f_i({\bf x},t)-f_i^{(0)}({\bf x},t)]
+ \delta h_i ,
\label{eq:12}
\end{equation}
 where the $h_i$ are chosen as:
\begin{equation}
 h_0 = 0, \;\;\;\; h_i = +G/4, \;\; i = 1,2,6,  \;\;\;\;
 h_i = -G/4, \;\; i = 3,4,5 ,
\label{eq:13}
\end{equation}
so that
\[
\sum_{i} h_i= 0, \;\;  \sum_{i} h_i {\bf e}_i = \rho {\bf g}, \;\;
\sum_{i} h_i e_{i\alpha} e_{i \beta} = 0 .    \]

Now suppose there exists a solution $f_i({\bf x}, t)$
of Eq.~(\ref{eq:12})  and it exactly represents
the Poiseuille flow. We then expect the following properties:

\noindent 1.  $f_i({\bf x}, t)$ is steady (independent of $t$). \\
2.  $f_i({\bf x}, t)$ is independent of $x$, hence is only a function
of $y$, denoted by $f_i(y)$. \\
3.  $f_2(y) = f_6(-y)$, and $f_3(y) = f_5(-y)$ from the  symmetry
of the flow. \\
4.  $\sum_i f_i(y) = \rho $ (constant).   \\
5.  $\sum_i f_i(y) e_{ix}  = \rho u(y)$ , where $u(y) = u_0(1-y^2)$
(remember $L = 1$). \\
6.  $\sum_i f_i(y) e_{iy}  = 0$ .

According to Eq.~(\ref{eq:3}), the equilibrium distributions
are given by
\begin{eqnarray}
 f_0^{(0)} = d_0 - \rho u^2, \;\;\; (u = u(y)) \mbox{\hspace{1in}}  \nonumber
\\
 f_1^{(0)} = d + \frac{\rho}{3} u + \frac{\rho}{2}  u^2 , \;\;
 f_4^{(0)} = d - \frac{\rho}{3} u + \frac{\rho}{2}  u^2 ,  \nonumber \\
 f_2^{(0)} = d + \frac{\rho}{6} u , \;\;
f_3^{(0)} = d - \frac{\rho}{6} u,  \mbox{\hspace{1in}}  \nonumber \\
 f_5^{(0)} = d - \frac{\rho}{6} u , \;\;
 f_6^{(0)} = d + \frac{\rho}{6} u .\mbox{\hspace{1in}}
\label{eq:14}
\end{eqnarray}
Using properties 1,2 and Eq.~(\ref{eq:12}) for $i=0$ gives:
\[
 f_0(y)  =  f_0(y) - \frac{1}{\tau} (f_0(y) - f_0^{(0)}(y)),  \]
hence
\begin{equation}
 f_0(y)  =   f_0^{(0)}(y) =  d_0 - \rho u^2 .
\label{eq:15}
\end{equation}
Eq.~(\ref{eq:12}) for $i=1$ gives:
\[
 f_1(y)  =  f_1(y) - \frac{1}{\tau} (f_1(y) - f_1^{(0)}(y)) + \delta G/4,  \]
hence
\begin{equation}
 f_1(y)  =   f_1^{(0)}(y) + \tau \delta G/4 .
\label{eq:16}
\end{equation}
Similarly
\begin{equation}
 f_4(y)  =   f_4^{(0)}(y) - \tau \delta G/4 .  \label{eq:17}
\end{equation}
It is seen that $f_0, f_1, f_2$ are functions of $y^4$, $y^2$ through
dependence of $u$ and $u^2$.
To find the remaining $f_i(y)$, we note that
$f_i^{(0)}, i = 2,3,5,6$ have no $u^2$ term and
thus  are functions of $y^2$ only, so the following
form is suggested:
\begin{equation}
 f_i(y)  =  a_i + b_i y + c_i y^2 , \;\; i = 2,3,5,6,
\label{eq:18}
\end{equation}
where  the twelve unknown coefficients
$a_i, b_i, c_i$ depend on flow quantities, $\tau$, $dy$, but not
on $y$.
Using property 3, we obtain
\begin{eqnarray}
  a_2 + b_2 y + c_2 y^2= a_6 - b_6 y + c_6 y^2 , \nonumber \;\;\;\;
  a_3 + b_3 y + c_3 y^2= a_5 - b_5 y + c_5 y^2 ,
\label{eq:19}
\end{eqnarray}
which should be true for any $y$, thus:
\begin{eqnarray}
  a_6 = a_2, \; b_6 = -b_2, \; c_6 =  c_2  \nonumber \\
  a_5 = a_3, \; b_5 = -b_3, \; c_5 =  c_3 .
\label{eq:20}
\end{eqnarray}
Similarly, using property 6, we find:
\begin{equation}
 2 b_2 y + 2 b_3 y = 0, \;\;\;\; \mbox{hence} \;\;\;\; b_3 = - b_2.
\label{eq:22}
\end{equation}
Property 4 with information obtained gives:
\begin{equation}
 2(a_2+a_3) + 2(c_2+c_3)y^2 + f_0(y) + f_1(y)  +  f_4(y) = \rho ,
\label{eq:23}
\end{equation}
On using the expressions for $f_0, f_1, f_4$ given in
Eq.~(\ref{eq:15} - \ref{eq:17}), we find
\begin{equation}
a_2+a_3 = \frac{1}{2} (\rho - d_0 -2d) = 4d \;\;
\mbox{(on using the expression of $d$)}, \;\;\;\;
 c_3 + c_2 = 0;
\label{eq:24}
\end{equation}
which gives
\begin{equation}
 c_3 =- c_2 , \;\;\;\; a_3 = 2d - a_2.
\label{eq:25}
\end{equation}
Similarly, property 5 with information obtained yields:
\begin{equation}
a_2 =\frac{1}{6}\rho u_0+d-\tau \delta G/4 ;
 \;\;\;\;
 c_2 =- \frac{1}{6} \rho u_0.
\label{eq:27}
\end{equation}
At this point, only $b_2$ remains unknown.  Using Eq.~(\ref{eq:12})
for $i=2$, we have
\begin{equation}
 f_2(y+dy)=f_2(y)-\frac{1}{\tau} (f_2(y) - f_2^{(0)}(y)) + \delta G/4 ,
\label{eq:28}
\end{equation}
where $dy$ is the vertical spacing between two lattice rows and
$dy = (\sqrt{3}/2)\delta$.
On using the expression  for $f_2^{(0)}$, we can obtain:
\begin{equation}
 c_2[y^2+2y dy+(dy)^2]+b_2y+b_2dy+a_2 =(1-\frac{1}{\tau})(c_2y^2+b_2y+a_2)
 +\frac{1}{\tau}[d+\frac{1}{6}\rho u_0(1-y^2)]+\delta G/4.
\label{eq:29}
\end{equation}
The balance of terms linear in $y$ yields:
\begin{equation}
 b_2 =  - 2\tau c_2 dy = \frac{1}{3} \tau \rho u_0 dy ,
\label{eq:30}
\end{equation}
and fortunately the equations for the coefficients of $y^2$ and $y^0$
are both satisfied.
It is easy to check that the evolution  equations for
$f_3, f_5, f_6$ are all satisfied with the choice of $a_i, b_i, c_i$
obtained so far.

Putting these results all together, we find that the quantities
\begin{eqnarray}
 f_0 = d_0 - \rho u^2, \;\;\;  \mbox{\hspace{2.1in}}  \nonumber \\
 f_1 = d + \frac{\rho}{3} u + \frac{\rho}{2}  u^2 +\tau \delta G/4,
\mbox{\hspace{1.3in}} \nonumber \\
 f_4 = d - \frac{\rho}{3} u + \frac{\rho}{2}  u^2 -\tau \delta G/4,
\mbox{\hspace{1.3in}} \nonumber \\
 f_2=-\frac{1}{6}\rho u_0 y^2+\frac{1}{3}\tau\rho u_0ydy+
      \frac{1}{6}\rho u_0+d-\tau\delta G/4,\nonumber \\
 f_3=+\frac{1}{6}\rho u_0 y^2-\frac{1}{3}\tau\rho u_0ydy-
      \frac{1}{6}\rho u_0+d+\tau\delta G/4, \nonumber \\
 f_5=+\frac{1}{6}\rho u_0 y^2+\frac{1}{3}\tau\rho u_0ydy-
      \frac{1}{6}\rho u_0+d+\tau\delta G/4, \nonumber \\
 f_6=-\frac{1}{6}\rho u_0 y^2-\frac{1}{3}\tau\rho u_0ydy+
      \frac{1}{6}\rho u_0+d-\tau\delta G/4,
\label{eq:31}
\end{eqnarray}
all satisfy properties 1-6 and that they together with the equilibrium
distribution
given in Eq.~(\ref{eq:14}) satisfy the LBGK equation, Eq.~(\ref{eq:11}).
Hence it is an exact representation of the Poiseuille flow.

Next, let us see to what boundary condition the solution in Eq.~(\ref{eq:31})
corresponds.  Taking the bottom boundary with $y=-1, u=0$, we have
\begin{eqnarray}
 f_0 = d_0 ,\;\;
 f_1 = d + \tau \delta G/4 , \;\;
 f_4 = d - \tau \delta G/4 \mbox{\hspace{0.5in}}  \nonumber \\
 f_2= -\frac{1}{3}\tau\rho u_0dy+d-\tau\delta G/4 , \;\;
 f_3= +\frac{1}{3}\tau\rho u_0dy +d+\tau\delta G/4 \nonumber \\
 f_5= -\frac{1}{3}\tau\rho u_0dy +d+\tau\delta G/4 , \;\;
 f_6= +\frac{1}{3}\tau\rho u_0dy+ d-\tau\delta G/4.
\label{eq:32}
\end{eqnarray}
It is seen that on the bottom, after the collision and forcing,
$f_2 = f_5 - 2 \tau\delta G/4$,
 $f_3 = f_6 + 2 \tau\delta G/4$. Hence,
if a bounce-back boundary condition $f_2 = f_5, f_3 = f_6, $ on
$f_2, f_3$ is applied at the bottom to replace the collision and forcing
step, the error is of order
$\delta$.  This shows that the bounce-back boundary condition
is first-order accurate.  This has been confirmed in computations
\cite{hou,zieg,qian4}.

To obtain the steady-state
analytical solution in the LBGK simulations,
the boundary
condition should be suitably chosen for the simulation.
The boundary condition proposed
by Nobel et al. \cite{nob} is a suitable choice.  If we are looking at
a node $B$ on the bottom, after streaming, $f_2$ and $f_3$ are empty at
the node $B$ since no particle is coming from outside.  Then
Eq.~(\ref{eq:2})
with $u=v=0$ are used to determine $\rho, f_2, f_3$.
Then the normal collision with force as given in Eq.~(\ref{eq:12})
is applied to $f_i$ on the boundaries.
Suppose that initially, we use uniform density $\rho_0$ and zero velocity
through the flow field; then we compute $f_i^{(0)}$ and set $f_i= f_i^{(0)}$
through the field. Since there are no pressure gradients, it is natural
that the density at each node remains constant $\rho_0$ (confirmed by
simulations). Therefore  Eq.~(\ref{eq:2}) can be used to find
the unique $f_2, f_3$ with the
correct
density and velocity, hence it is consistent with the evolution of
$f_2, f_3$ in
the analytical solution.
Simulation results indicate
that the numerical solution with this boundary condition
approaches the analytical solution as $t \rightarrow \infty$.

Next, let us consider a plain Couette flow, where the flow between two
parallel plates (corresponding to $y=0$ and $y=1$)
are driven by the constantly moving top plate with velocity
$u_0$. In this case,  the solution is given by:
\begin{equation}
 u = u_0 y, \;\; 0 \leq y \leq 1,  \;\;\;\; v = 0, \;\;\;\;
 \nabla p = 0 ,
\label{eq:34}
\end{equation}
with $\rho$ a constant and with no body force.  So the LBGK
model Eq.~(\ref{eq:4}) is used.
Using a similar procedure,
we find the analytical solution of Eq.~(\ref{eq:4}) representing
the Couette flow:
\begin{eqnarray}
 f_0 = d_0 - \rho u^2,  \;\;
 f_1 = d + \frac{\rho}{3} u + \frac{\rho}{2}  u^2, \;\;
 f_4 = d - \frac{\rho}{3} u + \frac{\rho}{2}  u^2,  \nonumber \\
 f_2=+\frac{1}{6}\rho u_0 y+d-\frac{1}{6}\tau \rho u_0dy, \;\;
 f_3=-\frac{1}{6}\rho u_0 y+d+\frac{1}{6}\tau \rho u_0dy, \nonumber \\
 f_5=-\frac{1}{6}\rho u_0 y+d+\frac{1}{6}\tau \rho u_0dy, \;\;
 f_6=+\frac{1}{6}\rho u_0 y+d-\frac{1}{6}\tau \rho u_0dy.
\label{eq:35}
\end{eqnarray}

We note that these analytical solutions are valid for any $u_0$, $\tau$
and $dy$.

\section{Analytical Solutions of the square lattice LBGK model}

 The square lattice Boltzmann BGK model is proven more robust than the
triangular model in
numerical simulations \cite{hou,sko}, it is important and interesting to find
analytical solutions for it.
The square lattice Boltzmann BGK model uses 3 types of particles.
 Particles of type 1 move along the x axis or the y axis
with speed
 ${\bf e}_{i} = (\cos(\pi(i-1)/2), \sin(\pi(i-1)/2), i = 1,2,3,4$,
and particles of type 2 move along the
diagonal directions with speed
${\bf e}_{i} = \sqrt{2} (\cos(\pi(i-4-\frac{1}{2})/2),
\sin(\pi(i-4-\frac{1}{2})/2), i = 5,6,7,8$.
Rest particles of type 0 with ${\bf e}_{0} = 0$
(speed zero) are also allowed at each node. Each node is connected to its
8 nearest neighbors by 8 links of length $\delta$ (in physical units) or
$\sqrt{2} \delta$ as shown in Fig.~2.
The  single-particle distribution function
$f_{i}({\bf x},t) $
again satisfies the LBGK model Eq.~(\ref{eq:4}) (with $i = 0,\cdots, 8$).
The density $\rho$ and the macroscopic velocoty ${\bf u}$ are still defined
in Eq.~(\ref{eq:2}).
For the square lattice,
the equilibrium distribution can be chosen in the following form for particles
of each type (the model d2q9 \cite{qian2}):
\begin{eqnarray}
f_{0}^{(0)}=\frac{4}{9}\rho[1-\frac{3}{2}{\bf u}\cdot{\bf u}],
\mbox{\hspace{1.65 in}}
\nonumber\\
f_{i}^{(0)}=\frac{1}{9}\rho[1+3({\bf e}_{i}\cdot
{\bf u})+\frac{9}{2}({\bf e}_{i}\cdot {\bf u})^2-\frac{3}{2}{\bf u}\cdot
{\bf u}],\;\; i =  1,2,3,4 \nonumber \\
f_{i}^{(0)}=\frac{1}{36}\rho[1+3({\bf e}_{i}\cdot
{\bf u})+\frac{9}{2}({\bf e}_{i}\cdot {\bf u})^2-\frac{3}{2}{\bf u}\cdot
{\bf u}], \;\; i =  5,6,7,8 ,
\label{eq:sq3}
\end{eqnarray}
with
${\displaystyle
\sum_{\sigma} \sum_{i} f_{\sigma i}^{(0)} =\rho, \;
 \sum_{\sigma} \sum_{i} f_{\sigma i}^{(0)} {\bf e}_{\sigma i} =\rho{\bf u}.}$
The macroscopic equations  of the model are the same as given in
Eqs.~(\ref{eq:5},\ref{eq:6}) with
${\displaystyle c_s^2= 1/3,}$
and
${\displaystyle \nu=\frac{2 \tau -1}{6} \delta}$.
To incorporate a body force in the model to model Poinseuille flow,
Eq.~(\ref{eq:12}) ( $i = 0, \cdots, 8$) is used,
with $h_i$ chosen in the following way \cite{qian1}:
\begin{eqnarray}
 h_{0} = 0, \ \  h_{1} = \frac{1}{3} G, \ \    h_{2} = 0,
 \ \ h_{3} = -\frac{1}{3} G, \ \  h_{4} = 0,  \nonumber \\
 h_{5} = h_{8} = \frac{1}{12} G, \ \ h_{6} = h_{7} = -\frac{1}{12} G,
\label{eq:sq13}
\end{eqnarray}

To derive an analytical solution of  Eq.~(\ref{eq:12}) for the square lattce,
we note that the six properties in Section 2 still applies except  that
property 3 is replaced by:  \\

3.  $f_2(y) = f_4(-y)$, $f_5(y) = f_8(-y)$ and $f_6(y) = f_7(-y)$
from the  symmetry of the flow. \\

Using a similar procedure as in Section 2, we can find that
\begin{equation}
 f_{0}(y)  =   f_{0}^{(0)}(y) =
\frac{4}{9} \rho (1 - \frac{3}{2} u^2),\;\; (u=u_0(1-y^2)),
\label{eq:sq15}
\end{equation}
\begin{equation}
 f_{1}(y)  =   \frac{1}{9} \rho (1 + 3u + 3 u^2) + \frac{2}{3}\tau
 \nu \rho u_0\delta .
\label{eq:sq16}
\end{equation}
\begin{equation}
 f_{3}(y)  =   \frac{1}{9} \rho (1 - 3u + 3 u^2) -  \frac{2}{3} \tau
 \nu \rho u_0 \delta.
\label{eq:sq17}
\end{equation}
and
\begin{equation}
 f_{i}(y)  =  a_{i} + b_{i} y + c_{i} y^2 +
  d_{i} y^3 + e_{i} y^4,  \;\; i = 2,4,5,6,7,8 ,
\label{eq:sq18}
\end{equation}
with
\begin{eqnarray}
  a_{4} = a_{2}, \; b_{4} = -b_{2}, \; c_{4} =  c_{2}, \;
  d_{4} = -d_{2}, \; e_{4} = -e_{2},
 \nonumber \\
  a_{8} = a_{5}, \; b_{8} = -b_{5}, \; c_{8} =  c_{5}, \;
  d_{8} = -d_{5}, \; e_{8} = -e_{5},
 \nonumber \\
  a_{7} = a_{6}, \; b_{7} = -b_{6}, \; c_{7} =  c_{6}, \;
  d_{7} = -d_{6}, \; e_{7} = -e_{6},
\label{eq:sq22}
\end{eqnarray}
and
\begin{eqnarray}
 a_{2} =  - 4\tau^4 \rho u_0^2 \delta^4+
  6\tau^3 \rho u_0^2 \delta^4 -
\frac{7}{3} \tau^2 \rho u_0^2 \delta^4 +
\frac{2}{3} \tau^2 \rho u_0^2 \delta^2 +
\frac{1}{6} \tau \rho u_0^2 \delta^4 -
\frac{1}{3} \tau \rho u_0^2 \delta^2 +
\frac{1}{9} \rho -
\frac{1}{6} \rho u_0^2 , \nonumber \\
 b_{2} =  4\tau^3 \rho u_0^2 \delta^3-
  4\tau^2 \rho u_0^2 \delta^3 +
\frac{2}{3} \tau \rho u_0^2 \delta^3 -
\frac{2}{3} \tau \rho u_0^2 \delta, \mbox{\hspace{2in}} \nonumber \\
 c_{2} = -2\tau^2 \rho u_0^2 \delta^2+
  \tau \rho u_0^2 \delta^2 +
\frac{1}{3} \rho u_0^2 , \;\;
 d_{2} = \frac{2}{3}\tau \rho u_0^2 \delta, \;\;
 e_{2} = -\frac{1}{6} \rho u_0^2 , \mbox{\hspace{1in}}
\label{eq:sq30}
\end{eqnarray}
\begin{eqnarray}
 a_{5} =  2\tau^4 \rho u_0^2 \delta^4-
  3\tau^3 \rho u_0^2 \delta^4 +
\frac{7}{6} \tau^2 \rho u_0^2 \delta^4 -
\frac{1}{3} \tau^2 \rho u_0^2 \delta^2 -
\frac{1}{6} \tau^2 \rho u_0 \delta^2 -
\frac{1}{12} \tau \rho u_0^2 \delta^4 + \nonumber \\
\frac{1}{6} \tau \rho u_0^2 \delta^2 +
\frac{1}{12} \tau \rho u_0 \delta^2  +
\frac{1}{36} \rho + \frac{1}{12} \rho u_0^2 +
\frac{1}{12} \rho u_0 + \frac{1}{6} \tau \nu \rho u_0  \delta ,
 \mbox{\hspace{1in}} \nonumber \\
 b_{5} =  -2\tau^3 \rho u_0^2 \delta^3+
  2\tau^2 \rho u_0^2 \delta^3 -
\frac{1}{3} \tau \rho u_0^2 \delta^3 +
\frac{1}{3} \tau \rho u_0^2 \delta +
\frac{1}{6} \tau \rho u_0 \delta ,\mbox{\hspace{1in}}  \nonumber \\
 c_{5} = \tau^2 \rho u_0^2 \delta^2-
  \frac{1}{2} \tau \rho u_0^2 \delta^2 -
\frac{1}{6} \rho u_0^2 -
\frac{1}{12} \rho u_0 , \;\;
 d_{5} = -\frac{1}{3}\tau \rho u_0^2 \delta, \;\;
 e_{5} = \frac{1}{12} \rho u_0^2 ,
\label{eq:sqf21}
\end{eqnarray}
and
\begin{eqnarray}
 a_{6} =  2\tau^4 \rho u_0^2 \delta^4-
  3\tau^3 \rho u_0^2 \delta^4 +
\frac{7}{6} \tau^2 \rho u_0^2 \delta^4 -
\frac{1}{3} \tau^2 \rho u_0^2 \delta^2 +
\frac{1}{6} \tau^2 \rho u_0 \delta^2 -
\frac{1}{12} \tau \rho u_0^2 \delta^4 + \nonumber \\
\frac{1}{6} \tau \rho u_0^2 \delta^2 -
\frac{1}{12} \tau \rho u_0 \delta^2  +
\frac{1}{36} \rho + \frac{1}{12} \rho u_0^2 -
\frac{1}{12} \rho u_0 - \frac{1}{6} \tau \nu \rho u_0  \delta ,
 \mbox{\hspace{1in}} \nonumber \\
 b_{6} =  -2\tau^3 \rho u_0^2 \delta^3+
  2\tau^2 \rho u_0^2 \delta^3 -
\frac{1}{3} \tau \rho u_0^2 \delta^3 +
\frac{1}{3} \tau \rho u_0^2 \delta -
\frac{1}{6} \tau \rho u_0 \delta , \mbox{\hspace{1in}} \nonumber \\
 c_{6} = \tau^2 \rho u_0^2 \delta^2-
  \frac{1}{2} \tau \rho u_0^2 \delta^2 -
\frac{1}{6} \rho u_0^2 +
\frac{1}{12} \rho u_0 , \;\;
 d_{6} = -\frac{1}{3}\tau \rho u_0^2 \delta, \;\;
 e_{6} = \frac{1}{12} \rho u_0^2 .
\label{eq:sqf22}
\end{eqnarray}
Eqs.~(\ref{eq:sq15},\ref{eq:sq16},\ref{eq:sq17}) together with
Eqs.~(\ref{eq:sq18},\ref{eq:sq22},\ref{eq:sq30},\ref{eq:sqf21},\ref{eq:sqf22})
 completely specify the analytical solution, which is a solution of
Eq.~(\ref{eq:12}) and it exactly represents the Poiseuille flow.

Next, let us see to what boundary condition this analytical
solution corresponds.  Taking the bottom boundary with $y=-1, u=0$, we find
the relation of $f_{\sigma i}$ after the collision and forcing:
\begin{eqnarray}
 f_{1} - f_{3} = \frac{4}{3} \tau\nu \rho u_0 \delta, \;\;
 f_{2} - f_{4} = -2 \delta^3 (4\tau^3\rho u_0^2-4\tau^2\rho u_0^2
+\frac{2}{3} \tau \rho u_0^2), \nonumber \\
 f_{5} - f_{7} = -\frac{2}{9} \tau^2\rho u_0 \delta^2
  +\frac{1}{9} \tau\rho u_0^2 \delta^2
+4\tau^3\rho u_0^2 \delta^3
-4\tau^2\rho u_0^2 \delta^3
+\frac{2}{3} \tau \rho u_0^2 \delta^3, \nonumber \\
 f_{6} - f_{8} = +\frac{2}{9} \tau^2\rho u_0 \delta^2
  -\frac{1}{9} \tau\rho u_0^2 \delta^2
+4\tau^3\rho u_0^2 \delta^3
-4\tau^2\rho u_0^2 \delta^3
+\frac{2}{3} \tau \rho u_0^2 \delta^3,
\label{eq:bc}
\end{eqnarray}
if a bounce-back boundary condition in which $f_{1}$ exchanges with $f_{3}$,
$f_{2} = f_{4}, $ $f_{5} = f_{7}, $ $f_{6} = f_{8}, $
is applied at the bottom to replace the collision and forcing
step, the error introduced to $f_{1}$ and $f_{3}$ is of order
$\delta$.  This shows that the bounce-back boundary condition
is first-order accurate.  This has been confirmed in computations
\cite{hou,zieg,qian4}.
To obtain the steady-state
analytical solution derived in this paper in LBGK simulations,
the boundary
condition should be suitably chosen.
No numerical simulation on a square lattice Boltzmann BGK model
has obtained an exact solution for the Poiseuille flow so far.
Of course,  if we use the analytical solution derived as the initial condition
and provide the right form of distribution functions on the boundaries,
we will be able to obtain the analytical solution in a simulation (confirmed
in simulations).
Specification of the analytical solution on boundary does not provide
a boundary condition of general purpose.
Nevertheless, the analytical solution will give some guidance
in developing  better boundary conditions of general purpose for the model.

The analytical solution for
 plain Couette flow are given by
\begin{eqnarray}
 f_{0} = \frac{4}{9}\rho (1-\frac{3}{2}u_0^2 y^2), \;\;
 f_{1} = \frac{1}{9}\rho (1+3u_0 y+3u_0^2 y^2), \;\;
 f_{3} = \frac{1}{9}\rho (1-3u_0 y+3u_0^2 y^2), \nonumber \\
 f_{2} = -\frac{1}{3}\tau^2\rho u_0^2\delta^2
 +\frac{1}{6}\tau\rho u_0^2\delta^2 +\frac{1}{9}\rho
 +\frac{1}{3}\tau\rho u_0^2\delta y
 -\frac{1}{6}\rho u_0^2 y^2 , \mbox{\hspace{1.5in}} \nonumber \\
 f_{4} = -\frac{1}{3}\tau^2\rho u_0^2\delta^2
 +\frac{1}{6}\tau\rho u_0^2\delta^2 +\frac{1}{9}\rho
 -\frac{1}{3}\tau\rho u_0^2\delta y
 -\frac{1}{6}\rho u_0^2 y^2 , \mbox{\hspace{1.5in}} \nonumber \\
 f_{5} = \frac{1}{6}\tau^2\rho u_0^2\delta^2
 -\frac{1}{12}\tau\rho u_0^2\delta^2
 -\frac{1}{12}\tau\rho u_0\delta +\frac{1}{36}\rho
 +(-\frac{1}{6}\tau\rho u_0^2\delta+ \frac{1}{12}\rho u_0)y
 +\frac{1}{12}\rho u_0^2 y^2 , \nonumber \\
 f_{6} = \frac{1}{6}\tau^2\rho u_0^2\delta^2
 -\frac{1}{12}\tau\rho u_0^2\delta^2
 +\frac{1}{12}\tau\rho u_0\delta +\frac{1}{36}\rho
 +(-\frac{1}{6}\tau\rho u_0^2\delta- \frac{1}{12}\rho u_0)y
 +\frac{1}{12}\rho u_0^2 y^2 , \nonumber \\
 f_{7} = \frac{1}{6}\tau^2\rho u_0^2\delta^2
 -\frac{1}{12}\tau\rho u_0^2\delta^2
 -\frac{1}{12}\tau\rho u_0\delta +\frac{1}{36}\rho
 +(+\frac{1}{6}\tau\rho u_0^2\delta- \frac{1}{12}\rho u_0)y
 +\frac{1}{12}\rho u_0^2 y^2 , \nonumber \\
 f_{8} = \frac{1}{6}\tau^2\rho u_0^2\delta^2
 -\frac{1}{12}\tau\rho u_0^2\delta^2
 +\frac{1}{12}\tau\rho u_0\delta +\frac{1}{36}\rho
 +(+\frac{1}{6}\tau\rho u_0^2\delta+ \frac{1}{12}\rho u_0) y
 +\frac{1}{12}\rho u_0^2 y^2 .
\label{eq:cou}
\end{eqnarray}
For the Couette flow, the top boundary is a moving  one, the analytical
solution given here will give a guidance in developing
a suitable boundary condition for moving boundaries.

We note that these analytical solutions are valid for any $u_0$, $\tau$
and $\delta$.
They will enhance our understanding of the method
and will give guidance in applications.

\vspace{1in}

{\bf Acknowledgments}

We  would like to thank
S. Chen,
D. Noble, G. McNamara, D. d'Humi\`{e}res,
D. Levermore, Y.H. Qian
for helpful discussions.  Q.Z. would like to thank
S. Chen for helping to arrange for his visit to
 the Los Alamos National Lab.

\vfill\eject

\vspace{1in}

{\bf Figure captions} \\

Figure 1. The geometry of the plain channel flow.

Figure 2. Schematic of a square lattice.

\end{document}